# Strain Effects on Band Structure and Dirac Nodal-Line Morphology of ZrSiSe

Bernardus Rendy[1, a)] and Eddwi H. Hasdeo[2,3, b)]

[1]*Engineering Physics Department, Faculty of Industrial Technology, Institut Teknologi Bandung, Indonesia*
[2]*Research Center for Physics, Indonesian Institute of Sciences, South Tangerang, Indonesia*
[3]*Department of Physics and Material Science, University of Luxembourg, Luxembourg*

[a)]Corresponding author: bernardusrendy@students.itb.ac.id
[b)]eddw001@lipi.go.id

**Abstract.** The Dirac nodal-line semimetals (DNLS) are new promising materials for technological applications due to its exotic properties, which originate from band structures dispersion and nodal-line behavior. We report a study on effects of several possibilities of strains in ZrSiSe DNLS on band structure dispersion and nodal-line behavior through the means of the density functional theory (DFT) calculations. We found that the Dirac nodal-line of ZrSiSe is robust to all strain with reasonable magnitude. Although, there are significant changes in gap, amplitude, and energy relative to Fermi energy. We also found the effective strains to tune the nodal-line and band structures are equi-biaxial tensile, uniaxial (100) tensile, and uniaxial (110) tensile strain.

## INTRODUCTION

In the last decade, discovery of a single-atom-thick honeycomb carbon layer – graphene – become one of main breakthroughs in science and technology [1-3]. The exotic electronic properties of graphene arise from massless nature of electrons depicted in two linearly dispersed bands that cross with each other at the Dirac node (or Dirac point). In three dimensions, this crossing node exists in discrete points or along closed curves in the Brillouin zone. The former belongs to the Dirac and Weyl semimetals while the latter is known as the nodal-line semimetals.

The Dirac nodal-line semimetals (DNLS) have attracted experimental and theoretical researches due to its exotic properties such as their topological nature [4], high charge carrier mobility [5], three-dimensional large butterfly magnetoresistance [6], and potential topological superconductivity [7]. The main feature of these non-symmorphic topological nodal-line semimetals is topologically protected linearly dispersing bands near the Fermi energy with continuous Dirac points along momentum-space, giving higher nodal-line density of states than that of other Dirac materials with discrete nodes [8]. This topological protection often comes from the non-symmorphic symmetries [9].

Inducing strain to tune properties of known materials are common practice, especially band structures such as in silicon [10]. Strain perturbations in nodal-line semimetals are interesting due to the Lifshitz transitions, tunability of optical properties [11-13], as well as its butterfly magnetoresistance [6]. Most common strains and pressure that has been studied in non-symmorphic nodal-line semimetals are uniaxial strain on axis perpendicular to the nodal-line ($z$ strain) and chemical pressure ($c/a$) [13-14]. The effect of strains in those studies are mostly in optical conductivity and plasmonic excitations [11-13]. However, to what extend does the nodal-line survive under applied strain remains to be understood [15].

In this study, we aim to understand the robustness of DNLS under several types of strains such as equi-biaxial tensile, equi-biaxial compression, hydrostatic, uniaxial (100), and uniaxial (110). We focus on ZrSiSe as a typical DNLS which slightly gapped due to spin-orbit coupling and its Dirac nodal-line shows sinusoidal oscillation along the closed loop in BZ. We found that the $k_z = 0$ nodal-line was robust through all previously mentioned strains under reasonable magnitude. However, the gap and amplitude of nodal-line changed. In several cases, the strains also introduce lowering of conduction bands, turning it into a topologically trivial metallic compound.

# METHOD

## Ab-Initio Calculation Details

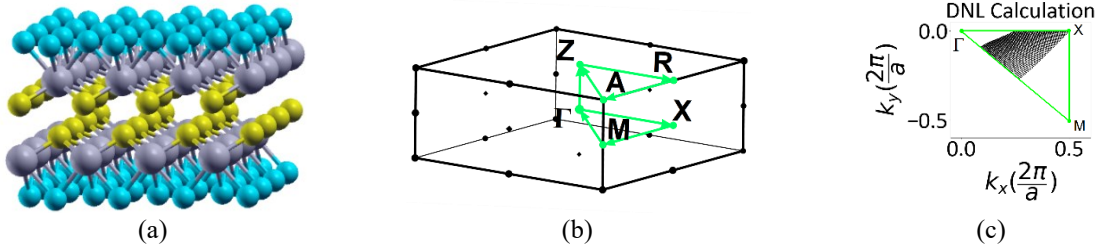

(a)        (b)        (c)

**FIGURE 1.** ZrSiSe (a) Schematic representation of structure. (b) *k*-path along First Brillouin Zone high symmetry points. (c) Schematic representation of nodal-line calculation at $k_z = 0$ in the First Brillouin Zone.

ZrSiSe is a layered crystal with a tetragonal structure and space group P4/*nmm* (No. 129). The structure of ZrSiSe is shown in Fig. 1a. The crystal graphics was generated using XCRYSDEN visualization package [16]. The plane-wave density functional theory (DFT) electronic structure calculation was implemented using QUANTUM ESPRESSO [17] simulation package with fully-relativistic optimized norm-conserving pseudopotentials [18-19]. To describe the exchange-correlation term, the generalized gradient approximation (GGA) [20] was applied in form of Perdew-Burke-Ernzerhof (PBE) correction. The Brodyen-Fletcher-Goldfarb-Shanno (BFGS) minimization scheme [21-22] was performed in structural optimization with force convergence criteria of $10^{-3}$ Ry/Bohr. The maximum plane-wave cut-off energy was taken as 80 Ry, while the electronic charge density was expanded in a basis cut off up to 320 Ry as optimized from previous study [23]. The Brillouin zone (BZ) integration was sampled with 24 x 24 x 8 *k*-points. Integration up to the Fermi surface was done with a smearing technique with a gaussian smearing of 0.002 Ry. The effect of spin-orbit coupling was taken into account in all of the calculation as it opens the ~40 meV gap at the Dirac points [23]. Self-consistency threshold for the total energy was set to $10^{-12}$ Ry.

## Strain, Band Structure, and Nodal-Line Calculations

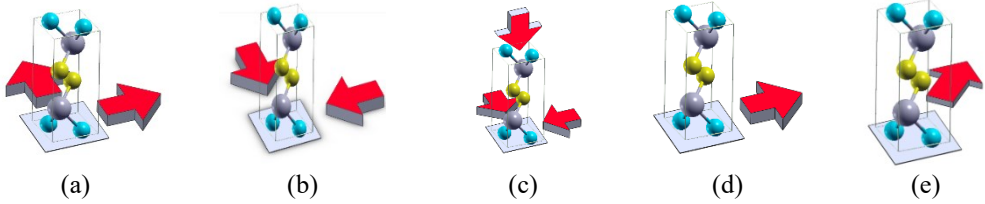

(a)       (b)       (c)       (d)       (e)

**FIGURE 2.** Schematic representation of (a) equi-biaxial tensile strain, (b) equi-biaxial compression strain, (c) hydrostatic strain, (d) uniaxial (100) tensile strain, and (e) uniaxial (110) tensile strain with angle displacement ($\Delta\gamma$) from 90°.

The calculated system lattice parameters were then varied according associated strain tensors under five categories: equi-biaxial compression strain, equi-biaxial tensile strain, hydrostatic strain, uniaxial (100) tensile strain, and uniaxial (110) tensile strain as shown in Fig. 2a – e, respectively. Associated stresses and pressures are linearly approximated from elastic constants in the previous study [24]. For every strained system, electronic band structure dispersion along high symmetry points in Fig. 1b and density of states were calculated. There are several nodal-lines in ZrSiSe, namely at $k_z = 0$, $\Gamma – X – R$, and $\Gamma – M – A$ planes. We have performed calculation on all those planes and found that all nodal-lines are gapped. Therefore, we focus on nodal-line at $k_z = 0$ plane. On the other hand, the $k_z = 0$ Dirac nodal-line (DNL) in this material is not located at the high symmetry points. In order to get the detailed picture of DNL, we scan the band structure along an array of 40 points from X (0.5,0,0) to a point between $\Gamma – M$ (-0.25,-0.25,0) each with 40 points parallel with line between $\Gamma$ and the points originated from the first array resulting in Fig. 1c where black dots are points of calculations with the assumption of mirror symmetry with the X – $\Gamma$ – X and M – $\Gamma$ – M plane. All of the band structures along high symmetry points and band structures near the nodal-lines are plotted with energy relative to the fermi energy (E-$E_f$).

# RESULT AND DISCUSSION

## Equilibrium State without Strain

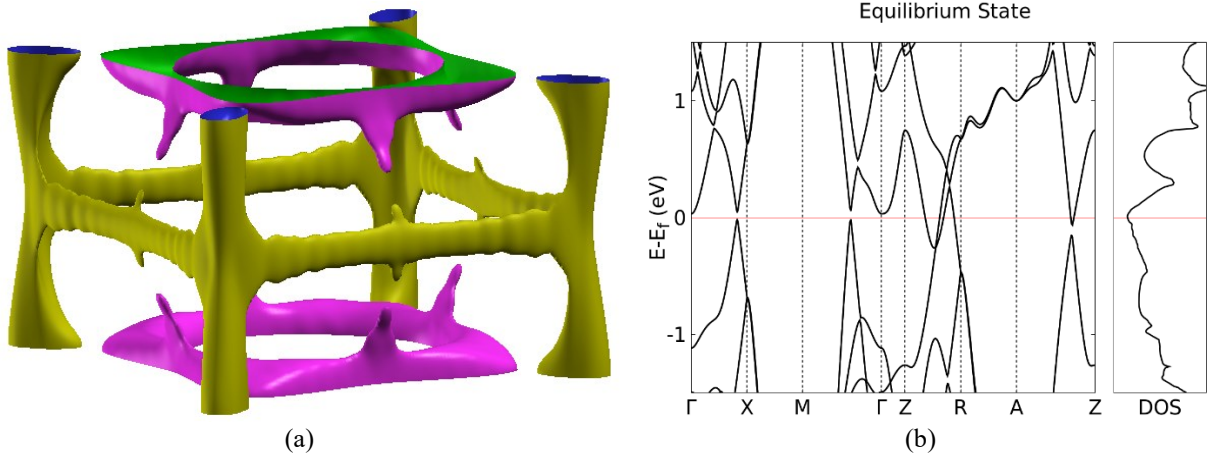

**FIGURE 3.** Equilibrium state (a) Calculated Fermi Surface. (b) Calculated band structure along high symmetry points.

Equilibrium state without strain exhibit a relaxed tetragonal structure with symmetry P4/*nmm* (No. 129). The Fermi surface under this condition gives a continuous cage-like surface with gap between Z-R and in between the $k_z = 0$ nodal-line and $k_z = \pm\pi/c$, tilted ~22° relative to $k_z$ as shown in Fig. 3a. The band structure dispersion along high symmetry points as shown in Fig. 3b displays ~40 meV gap due to spin-orbit coupling (SOC) with four-fold degeneracy in the DNL in which there are two valence bands and two conduction bands in each pair. The cage-like nodal-lines are protected by the non-symmorphic symmetries, but gapped out against SOC.

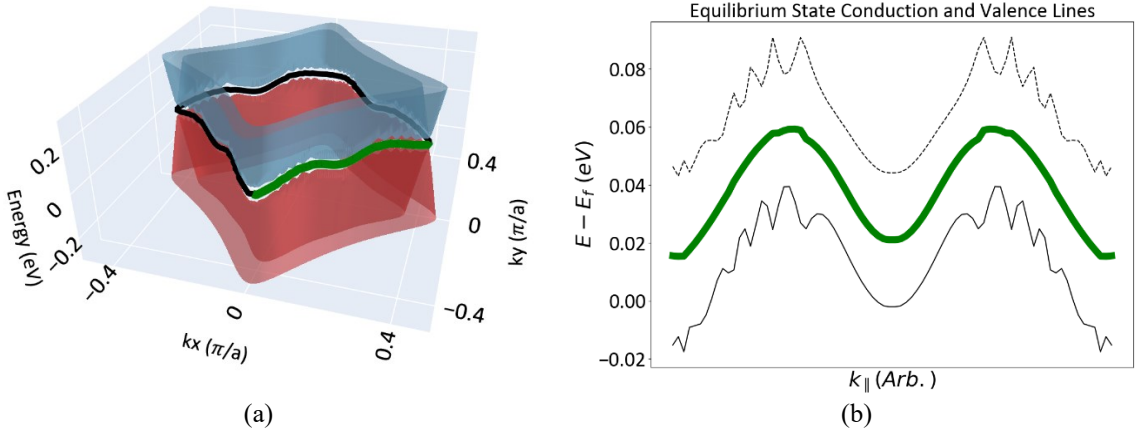

**FIGURE 4.** (a) Nodal-line and continuous Dirac cones at $k_z = 0$ in Brillouin Zone. (b) Equilibrium conduction (dashed lines) and valence (solid lines) bands at $k_z = 0$. $k_\parallel$ is the projection of green line to $k_x - k_y$ plane.

The nodal-line is located at wavevector defined by $k_\parallel$ at $k_x - k_y$ plane ($k_z = 0$) as shown in Fig. 4a. One can take a segment of nodal-line (the green line in Fig. 4a) representing the whole NL in the Brillouin zone. The energy along this nodal-line segment shows a sinusoidal feature with peak to trough differs by ~ 40meV (see Fig. 4b). The green band is the imaginary line in the middle of the valence (solid line) and conduction (dashed line) bands located around 39 meV above Fermi energy. There are linearly dispersing conduction and valence bands along this nodal-line where the linear characteristic is maintained for a range of ~1 eV from the lowest energy of the linear valence to highest energy of the linear conduction bands. The gap along the nodal-line varies from 39 to 66 meV with average of 50 meV. The conduction band always empty (lies above the Fermi energy) while valence band is partially filled. In the next section, we will show how the morphology of this NL and global band structures are deformed by strain.

## Equi-biaxial Tensile Strain

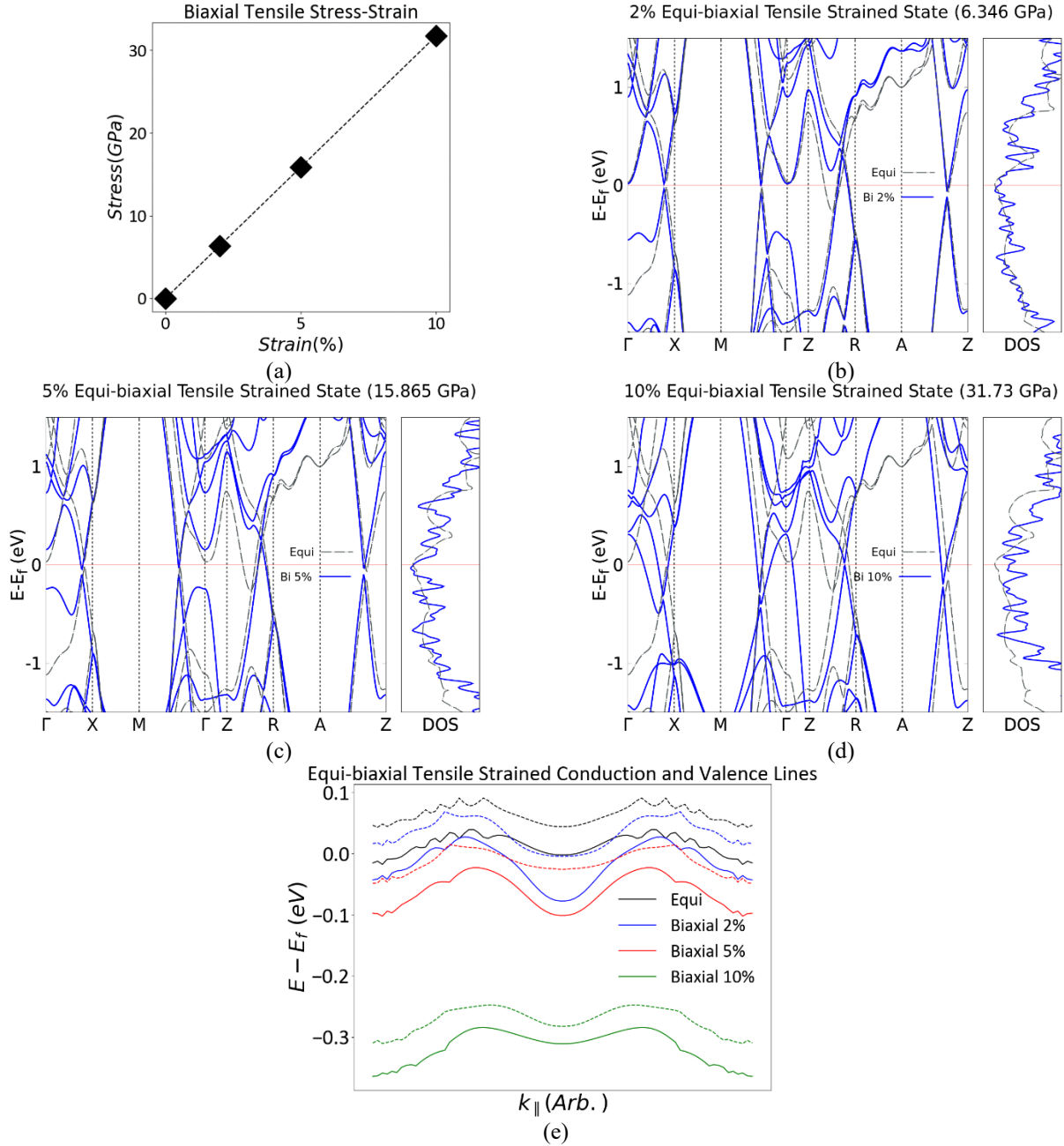

**FIGURE 5.** Equi-biaxial tensile strained (a) Linearly approximated stress due to corresponding strain in *x* and *y* direction. Resulting band structures and density of states for (b) 2%, (c) 5%, and (d) 10% equi-biaxial tensile strain. (e) Conduction bands (dashed lines) and valence bands (solid lines) of (b), (c), and (d) along $k_\parallel$.

In the equi-biaxial strains, we strained the cell on the *x* and *y* direction, leaving the *z* direction and atoms positions to be relaxed, giving minimum force and pressure. First, we begin by discussing the linearly approximated and interpolated stress-strain curve as shown in Fig. 5a. The magnitude of the stress required to achieve such strain is relatively large at the scale of 6-32 GPa for 2-10% strain. We note that these results are applicable in x and y axes symmetrically. The change in band structure dispersions along high symmetry points are as shown in Fig. 5b, 5c, and 5d. There is a transition of conduction bands in between Z – R moving from below to above Fermi energy. The gap near the Γ-point is also reduced with the increase of the valence band energy. Nevertheless, the valence and conduction band never coalesce. The trend from 2% to 10% shows that with increasing equi-biaxial tensile strain, the bandwidth of linear Dirac cone is reduced with nearly flattened out valence part of the cone in 10%. The change of density of states near the Fermi energy are not significant in 2% and 5%, but increased nearly 50% for 10% strain with also significant increase both below and above Fermi energy.

We also analyze the DNL near the Fermi energy by plotting the conduction bands and valence bands along $k_\parallel$ at $k_z = 0$. The equi-biaxial tensile strain reduces the global energy dispersion while keeping similar sinusoidal profile as shown in Fig. 5e. For the 2% tensile strain, gap varies from 33 to 73 meV with average of 49 meV. For the 5% tensile strain, the gap varies from 32 to 75 meV with average of 47.5 meV. Lastly, for the 10%, gap varies from 28 to 63 meV with average of 43 meV. There might be opportunity of adjusting the band structure dispersion, nodal-line gap and wave-like amplitude, and Fermi energy using strain by also considering relatively large strain, especially in the 5% and 10%. This result also suggests an alternative way to tune the DNL average energy relative to the Fermi energy while also tuning the conduction bands so that only DNL are present near the Fermi energy. We also need to consider that too much equi-biaxial tensile strain will increase the valence bands near Γ-point energy as demonstrated in 10% strain where the near Fermi energy bands are dominated by the valence bands near Γ-point, too much strain will also lead to increase of Fermi energy such that all of the DNL valence bands are filled and conduction bands are partially filled or even completely filled. The gap which increases near the M-point (~0.075 eV) might change the behavior of the system into an insulator rather than semimetal.

## Equi-biaxial Compression Strain

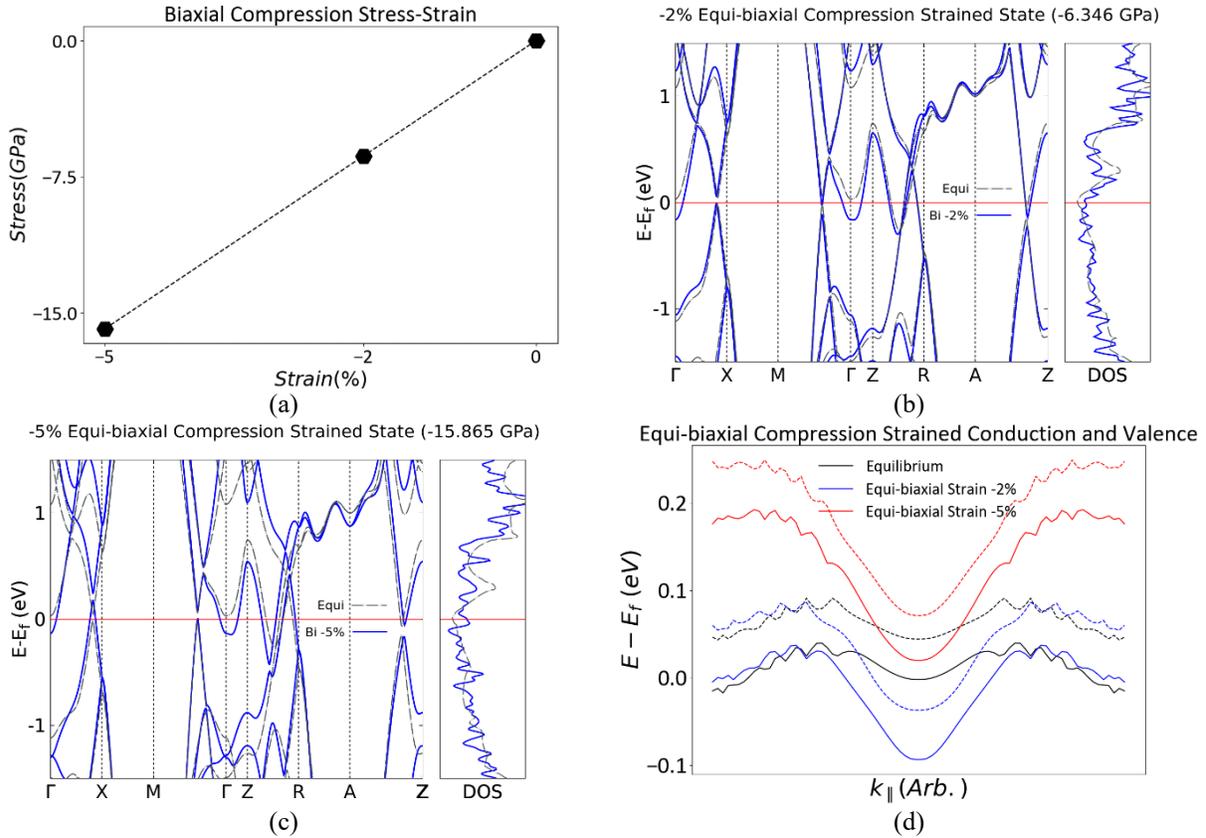

**FIGURE 6.** Equi-biaxial compression strained (a) Linearly approximated stress due to corresponding strain in *x* and *y* direction. Resulting band structures and density of states for (b) -2%, (c) -5% equi-biaxial compression strain. (d) Conduction bands (dashed lines) and valence bands (solid lines) of (b) and (c) along $k_\parallel$.

The stress-strain of biaxial compression strained is shown in Fig. 6a. The conduction band of the equi-biaxial compression strain structure slightly drops near the Γ-point to below the Fermi energy, creating an accessible conduction band to electronic ground state as shown in Fig. 6b and 6c, making the system more conductive. Further increase of compression to -5% lowers the conduction band even more. The density of states increased slightly in near the Fermi energy due to the lowered conduction bands and the change of Fermi energy relative to the DNL.

There are two-step changes in the $k_z = 0$ nodal-line. Firstly, small compression increases the amplitude of sinusoidal nodal-line energy. Secondly, further compression will elevate the valence band from full to partial occupation as shown in Fig. 6d. The DNL sinusoidal profile frequency is reduced from two full wave in equilibrium to half wave in -5% compression. The gap does not change significantly from -2% to -5% but change slightly from equilibrium to -2%. The -2% gap ranges from 35 to 68 meV with average of 50 meV. The result in -5% suggests that the compression will ascend the nodal-line at $k_z = 0$ with relatively large energy difference (~0.2 eV) between Dirac node minimum and maximum in the $k$-space. Note that these changes are only achievable under very large compression stress.

## Uniaxial (100) Tensile Strain

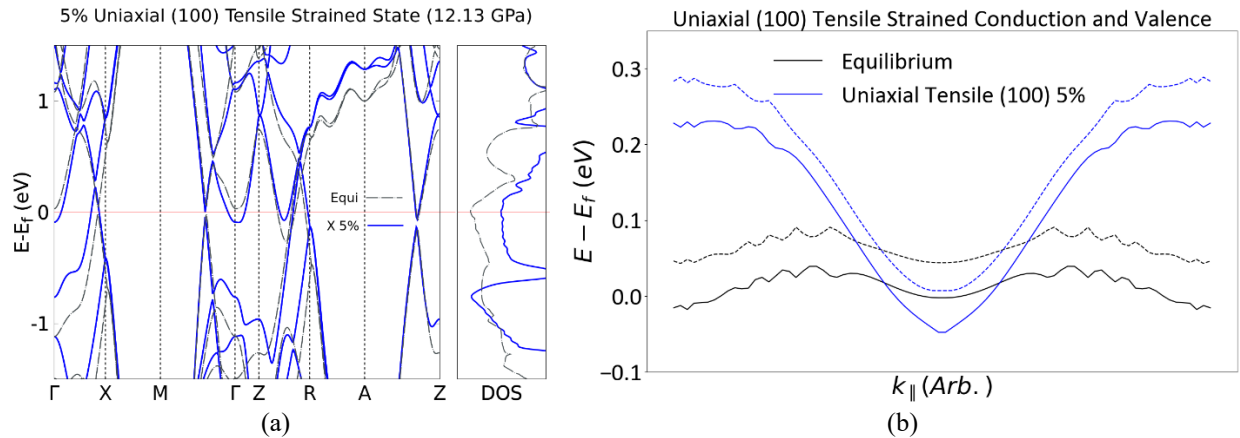

**FIGURE 7.** (a) Resulting band structures and density of states for 5% uniaxial (100) tensile strain. (b) Conduction bands (dashed lines) and valence bands (solid lines) of (a) along $k_\parallel$.

The uniaxial (100) tensile strain breaks the tetragonal unit cell, creating orthorhombic unit cell with less symmetry. The resulting band structure has combinations of tensile and compression effects as shown in Fig. 7a which are due to expansion of the cell in $x$ direction and compression in $y$ and $z$ direction. Conduction band near Γ-points moved from above to below Fermi energy while conduction bands between Z – R moved slightly higher to near Fermi energy. The density of states near the Fermi energy increased dramatically, but also followed by increase above and below the Fermi energy. The band structure between Z – R suggests overlap of linear valence band with conduction band, creating a half-closed gap in the $k_z = \pm\pi/c$ nodal-line. There is also reduced frequency of the sinusoidal nodal-line energy along $k_\parallel$ similar to the equi-biaxial compression but with higher amplitude up to ~280 meV. The gap in the nodal-line remain the same as equilibrium state around 30 and 66 meV with average of 43 meV as shown in Fig. 7b. There is also a huge difference (~0.25 eV) between minimum and maximum of the nodal-line. Stress needed in $x$ direction is 12.13 GPa, which is large but smaller than equi-biaxial strains for more changes in nodal-line.

## Hydrostatic Strain

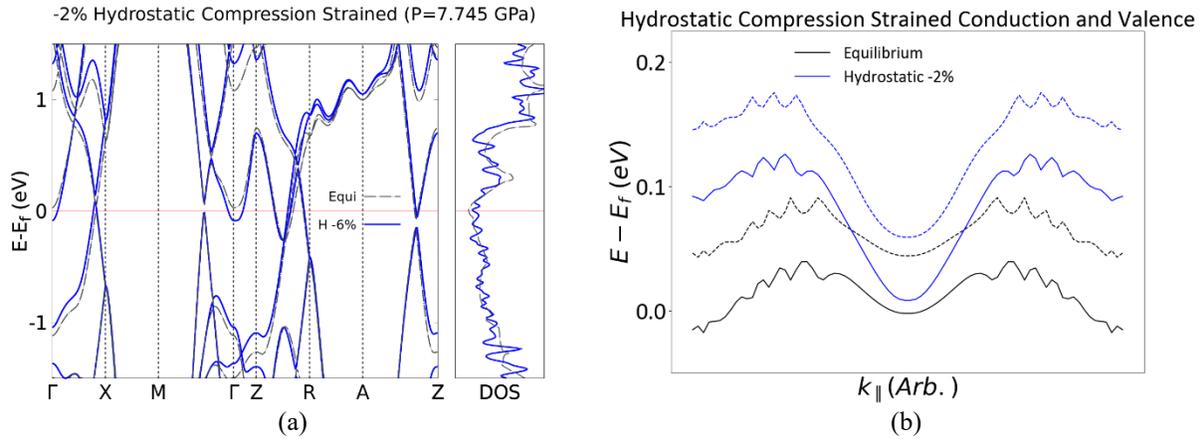

**FIGURE 8.** (a) Resulting band structure and density of state for -2% hydrostatic strain. (b) Conduction bands (dashed lines) and valence bands (solid lines) of (a) along $k_\parallel$.

Hydrostatic or equi-triaxial compression strain is achieved by inducing -2% strain for each cell parameters and letting the atoms relax to achieve minimum energy. The effect on band structure is not too significant, similar to the effect of biaxial compression with no significant increase on density of states as shown in Fig. 8a. Zooming into the DNL structure, all of the nodal-line bands rise above Fermi energy with significant increase in difference between maximum and minimum energy of the Dirac node (~125 meV), with the gap ranges from 36 to 64 meV and average of 48 meV as shown in Fig. 8b. The result suggest that correct amount hydrostatic strain can be used to analyze the effect of nodal-line lifting without changing the band structure along high symmetry points.

## Uniaxial (110) Tensile Strain

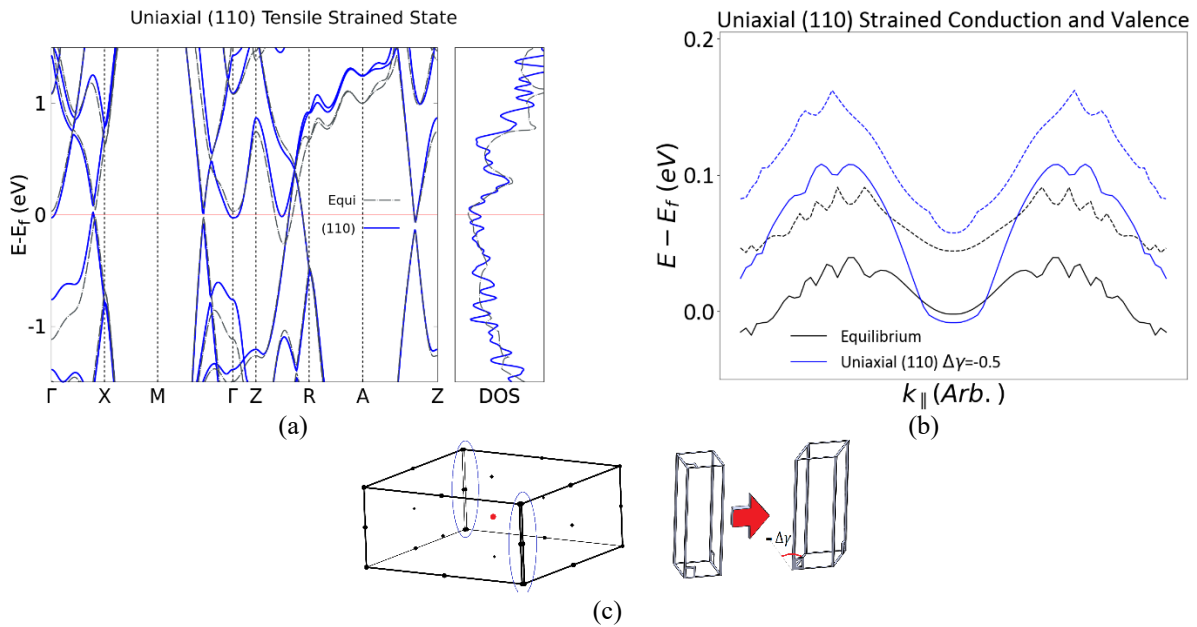

**FIGURE 9.** (a) Resulting band structure and density of states (b) Conduction bands (dashed lines) and valence bands (solid lines) along $k_\parallel$. (c) Changes of Brillouin Zone and unit cell to monoclinic.

Uniaxial strain (110) is a unique strain due to its condition that break tetragonal unit cell, changing the unit cell into monoclinic with even lesser symmetry than orthorombic as shown in Fig. 9c. The uniaxial strain (110) introduces change in γ or the angle between *x* and *y*-axis. In our simulation, we take small angle displacement (Δγ = -0.5°) as it is practically possible. The Dirac cone between Z – R ($k_z = \pm \pi/c$) was lifted in energy and tilted such that it overlaps with existing conduction band creating higher degeneracy but with less bandwidth and less gap where the gap half-closed by other conduction band near Z – R as shown in Fig. 9a. The resulting DNL is calculated in the new Brillouin Zone which is a monoclinic Brillouin Zone as shown in Fig. 9c. The strain changes the nodal-line significantly while still maintaining the sinusoidal structure with increased amplitude and average energy as shown in Fig. 9b. The gap ranges from 34 to 76 meV with average of 50 meV. There is also relatively large energy difference (~100 meV) between maximum and minimum of the Dirac nodes.

### Dirac Nodal-Line Morphology and Average Energy

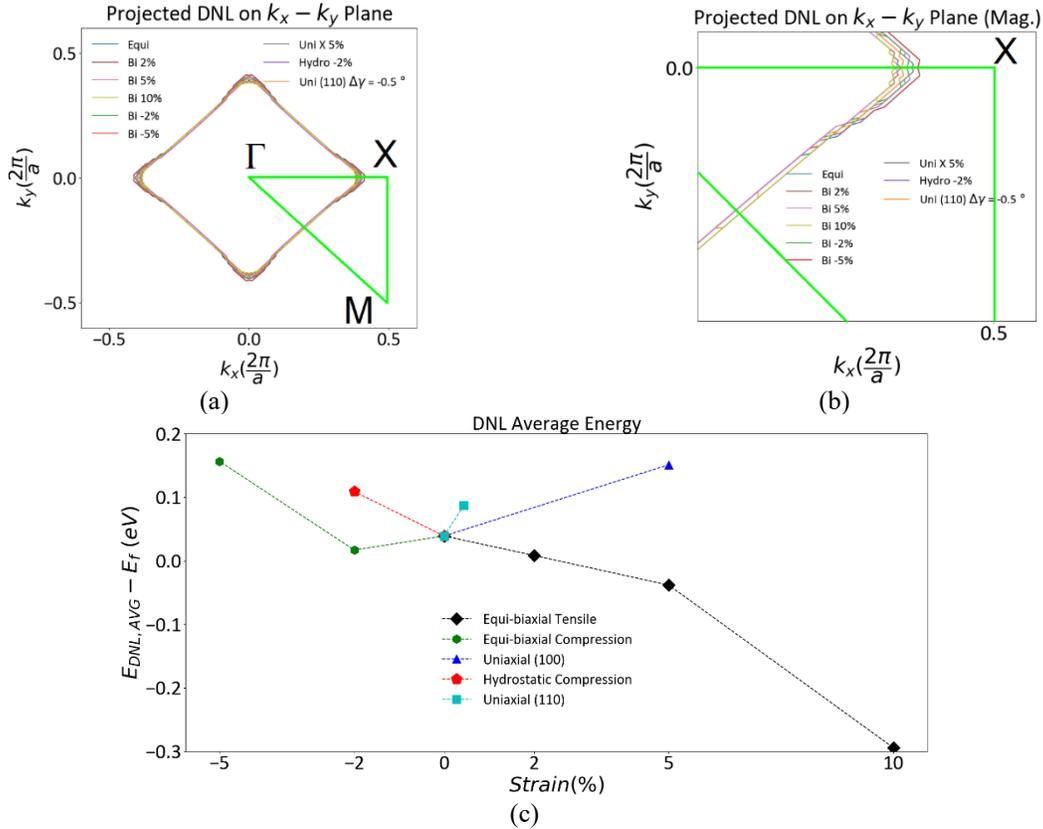

**FIGURE 10.** (a) Projected DNL on $k_x - k_y$ plane. (b) Near high symmetry points magnification of (a). Equilibrium DNL projection overlapped with biaxial -5% strained DNL projection. (c) DNL average energy due to various strain.

The DNL average energy is defined as average energy of the imaginary line between conduction and valence bands in DNL relative to the Fermi energy. DNL average energy changes due to strain can be seen in Fig. 10c. DNL average energy is reduced when equi-biaxial tensile strain is applied and increased when hydrostatic compression, uniaxial (100) tensile, and uniaxial (110) tensile strain are applied. Equi-biaxial compression reduce the DNL average energy for small magnitude and increase it for large magnitude. The most efficient strains in tuning DNL average energy are equi-biaxial tensile, uniaxial (110) tensile, and uniaxial (100) tensile strain. Apart from analyzing nodal-line energy, we also analyze the DNL morphology on $k_x - k_y$ plane as shown in Fig. 10a. All strained systems experience minor reduction in cross-sectional area. The 10% equi-biaxially tensile strained systems exhibited maximum reduction in cross-sectional area compared to equi-biaxial compression, uniaxial (100), hydrostatic, and uniaxial (110) strains as shown in figure 10(b). It is arresting that nodal-line morphology remains robust under applied strain.

# CONCLUSION

Various strain in different magnitudes and directions have been studied in gapped nodal-line of ZrSiSe by the means of density functional theory (DFT). Equi-biaxial tensile tends to reduce DNL energy relative to the Fermi energy, while equi-biaxial compression, uniaxial (100) tensile, hydrostatic compression, and uniaxial (110) tensile strain tends to increase it. The nodal-line gap average relatively stays the same except in uniaxial (100) 5% tensile strain where it decreased by 14%. These nodal-line changes are easiest to be observed in equi-biaxial tensile strain, uniaxial (100) tensile strain, and uniaxial (110) tensile strain considering stress and pressure magnitude. Finally, we conclude, under the definition of nodal-line as a continuous Dirac node over $k$-space, that the ZrSiSe nodal-line is robust against previously mentioned types and magnitudes of strain with different energy deviation to the Fermi energy and difference between maximum and minimum Dirac node. All strains might induce electronic topological transition (ETT) near Γ – points while equi-biaxial tensile, uniaxial (100), and uniaxial (110) might induce ETT near Z – R points. No degeneracy lifting occurs in all strains.

# ACKNOWLEDGEMENTS

The computation in this work has been done using the facilities of HPC LIPI, Indonesian Institute of Sciences (LIPI) and HPC facilities of the University of Luxembourg [25] (see http://hpc.uni.lu). EHH acknowledges ATTRACT 7556175 and CORE 11352881.